\documentstyle[12pt, epsf]{article}
\def\Lao{\lambda_{\alpha}^{obs}}
\def\Lbo{\lambda_{\beta}^{obs}}
\def\Lal{\lambda_{\alpha}^{lab}}
\def\Lbl{\lambda_{\beta}^{lab}}
\def\dLo{\lambda_{\alpha}^{obs}/a-\lambda_{\beta}^{obs}/b}
\def\dLl{\lambda_{\alpha}^{lab}/a-\lambda_{\beta}^{lab}/b}

\begin{document}

\begin{flushright} USM-TH-83 \end{flushright}
\vspace*{.1in}

\begin{center} \Large\bf Extra Dimensions in the Early Universe \end{center}

\begin{center}
Douglas J. Buettner\footnote[1]{Douglas\_Buettner@xontech.com}
\\XonTech Corporation\\ 6862 Hayvenhurst Avenue, Van Nuys, CA
91406 \end{center}
\vspace*{.10in}
\begin{center}
P.D. Morley\footnote[2]{pmorley@eisi.com}
\\EIS International \\ 555 Herndon Parkway, Herndon, VA 20170
\end{center}
\vspace*{.10in}
\begin{center}
Ivan Schmidt\footnote[3]{ischmidt@fis.utfsm.cl. Work supported in
part by Fondecyt (Chile) grant 1980149, and by a C\'atedra
Presidencial (Chile).} \\ Department of Physics, Universidad
T\'ecnica Federico Santa Mar\'\i a \\
Casilla 110-V, Valpara\'\i so, Chile
\end{center}
\vspace*{.10in}
\begin{abstract}
\baselineskip=0.7cm
We investigate the possible occurrence of extra spatial dimensions
($D = 3+\epsilon$) in the early
universe. A detailed calculation is presented which shows that the
crucial signal is the apparent inequality of the cosmological
$Z$-term between
matching Lyman alpha (Ly$_{\alpha}$) and Lyman beta (Ly$_{\beta}$)
spectral lines, both emission and
absorption, when using the present epoch (laboratory) wavelengths. 
We present 
preliminary upper limits to the value of epsilon, to be improved by
direct, more careful analysis of the spectra.
We take catalogued quasar Ly$_{\alpha}$ forest data and perform
Student's t-test to determine whether we should reject the null hypothesis
(no fractal dimensions). Finally, a $\chi^{2}$ analysis is done for
fitting $\epsilon$ in the early universe. The statistical tests and experimental data
are all consistent with $\epsilon = 0$ for $Z \leq 3.3$, but the 
experimental data support non-zero $\epsilon$ values for $Z \geq 
4$. However, it should be emphasized that the non-zero values of
epsilon found for $Z \geq 4$ may be due to undiscovered systematic errors 
in the original data. 
\end{abstract}

\vspace{.1in}

\vspace{.2in}
(To be published in Physics Letters B.)
\newpage
\parskip=9pt

Many modern physical theories predict \cite{KK} that space has
more than 3 spatial dimensions, some of which would reveal
themselves only at small distances. This means that because of an
expanding universe, the effective dimension of space may be a
time-dependent parameter. In fact, the dimension of space is an
experimental quantity \cite{AZ}, whose present epoch value and
past distant value may be different.
Since we are interested in the possibility that the early
universe had fractal ($D = 3+\epsilon$ with $\epsilon \ll 1$)
spatial dimensions, the natural place for investigation is in the
spectra of distant ($Z = \frac{\Delta \lambda}{\lambda}> 2.5$) quasars.
In this regard, the quasar's Lyman spectra and the 
existence of spectroscopic Ly$_{\alpha}$ forests \cite{MR}
provide an ideal opportunity to probe the fractal nature of
early universe space on an atomic length scale, provided we can
identify a suitable experimental signal.

Past researchers have investigated the possibility of a
time-varying fine structure constant ($\alpha$), using the
hyperfine multiplets of absorption lines in quasar ionized iron
and magnesium spectra, and obtaining an upper bound of $\delta
\alpha / \alpha < 1.1 \times 10^{-5}$ \cite{JKW}. In this paper
we use ancient quasar light to probe the dimension of space in
the early universe, specifically
matching (same $Z$) Ly$_{\alpha}$ and Ly$_{\beta}$ hydrogen
lines. As far as we know, both the main idea of this paper and 
the specific matching lines technique have not been considered
before. Therefore we discuss in detail the errors involved, which
are both experimental and theoretical.

To lowest order in quantum corrections, the atomic electromagnetic potential
is the solution of the Poisson equation, whose form depends on the
spatial dimension\footnote[4]{For example, the 2-dimensional
Coulomb solution is $V =\ln (\rho/ \rho_{o})$.}
In \cite{BM} the D(spatial)-dimensional
Schr\"odinger equation is derived. From a mathematical
point-of-view \cite{DH}, expectation values in quantum mechanics
can be analytically continued into continuous dimensions and it
becomes meaningful to construct a Taylor's expansion for $D =
3+\epsilon$ with $\epsilon \ll 1$

\begin{equation}
<|H|>\big|_{D=3+ \epsilon} =<|H|>\big|_{D=3} + \frac{d <|H|> }{d
D}\big|_{D=3}\epsilon + \cdots  \; .
\end{equation}

Reference \cite{BM} has proven the generalized Hellmann-Feynman
theorem

\begin{equation}
\frac{d <|H|> }{d D}\big|_{D=3}=<| \frac{\partial H}{\partial
D}\big|_{D=3} |> \end{equation}

where $H$ is the D-dimensional Hamiltonian.

An interesting aspect of fractal dimensions is that the
electric charge $e$ becomes a dimensional constant. The scaling
is $e \sim l^{(D-3)/2}_{o}$. Thus a length parameter $l_{o}$
enters into the problem. In discussing the present-epoch Lamb
shift, reference \cite{BM} has shown that if this length
parameter is not very much smaller than the Planck length
($10^{-33}$ cm), then it makes a negligible contribution to
atomic energy levels.

Using eqn. (2) we have calculated the shift in energy levels due
to the first order $\epsilon$ contributions to atomic hydrogen,
and these are given in Table 1. The atomic energy levels are

\begin{equation} E(nl) = E(nl)|_{laboratory} + \Delta E(nl)
\times \epsilon \; . \end{equation}

 From these energy levels we obtain the Ly$_{\alpha}$ ($\alpha$)
and Ly$_{\beta}$ ($\beta$) rest frame transition wavelengths

\begin{equation}
\lambda _{\alpha }=\lambda _{\alpha }^{lab}+a\,\epsilon, \quad
\; {\rm and}\quad \;
\lambda _{\beta }=\lambda _{\beta }^{lab}+b\,\epsilon
 \; , \end{equation}

where $\lambda_{\alpha}^{lab} = 1215.67 \mbox{\normalsize\/ \AA}$ and
$\lambda_{\beta}^{lab} = 1025.72 \mbox{\normalsize\/ \AA}$ are the
laboratory (present epoch) wavelengths, and
$a=1418.27\mbox{\normalsize\/ \AA}$ and $b=1111.18
\mbox{\normalsize\/ \AA}$ are the corresponding shifts
due to the extra dimensions. Equation (4) demonstrates that a
fractal dimension of space affects the two primary cosmological
transitions differently. From the physics viewpoint, the kinetic energy and centripetal
terms in the D-dimensional Schr\"odinger equation are most 
sensitive to the presence of fractal dimensions and make the 
atomic energy levels ideal indicators of additional space-time 
dimensions.

\begin{table}      \begin{tabular}{lc}
     \multicolumn{2}{c}{epsilon contributions} \\  \hline
     state vector $|nl(j)>$  & $\Delta E(nl)$ \\
     \mbox{} & \mbox{} \\
     $|1s(1/2)>$  &  1/2 \\
     $|2s(1/2)>$  &  1/16  \\
     $|2p(1/2)>$ and $|2p(3/2)>$ &  1/16 \\
     $|3p(1/2)>$ and $|3p(3/2)>$ &  1/54 \\
     Lyman alpha & -7/16 \\
     Lyman beta & -13/27
     \end{tabular}
     \caption{$\epsilon$ contributions in units of
     $\alpha^{2} \mu c^{2} = 27.19658$ eV, where $\mu$ is the reduced electron's mass.}
     \end{table}

The effect of a non-zero fractal dimension is as follows. When measurements
are taken of matching (same hydrogen cloud) Ly$_{\alpha}$ and
Ly$_{\beta}$ lines (both absorption and emission), it will
not be possible to obtain the same $Z$ shift using present epoch (i.e.
laboratory) rest wavelengths. Conversely, since matching
Ly$_{\alpha}$ and Ly$_{\beta}$
lines must have the same cosmological $Z$, the measurement of the
two red-shifted lines allows us to obtain the original early universe rest
frame frequencies, and determine statistically whether early universe
fractal dimensions exist. By equating $Z$ and $\epsilon$ for the two
transitions, we obtain two equations in the two unknowns $\lambda_{\alpha}$,
$\lambda_{\beta}$ which are the original early universe rest frame
transitions. From either solution, using eqns. (4), $\epsilon$ can
be deduced:

\begin{equation}
\epsilon = \frac{\lambda_{\alpha}^{obs}}{a} \frac{[
  \frac{\lambda_{\alpha}^{lab}}{a}
  - \frac{\lambda_{\beta}^{lab}}{b}]}
  {[
  \frac{\lambda_{\alpha}^{obs}}{a}
  - \frac{\lambda_{\beta}^{obs}}{b}
  ]} - \frac{\lambda_{\alpha}^{lab}}{a}
      \end{equation}
where $\lambda_{\alpha}^{obs}$, $\lambda_{\beta}^{obs}$ are the
respective Ly$_{\alpha}$, Ly$_{\beta}$ measured red-shifted
transition wavelengths, and $\lambda_{\alpha}^{lab}$,
$\lambda_{\beta}^{lab}$ are the laboratory measured wavelengths.

In order to obtain the unknown
quasar rest frame transitions from the measured cosmological red shifted lines, one
must have strict equality of the $Z$ factors. By choosing matching Ly$_{\alpha}$ and
Ly$_{\beta}$ lines, this is assured. Other spectra, such as metal ions, have narrower
lines than the hydrogen Lyman, but finding two matching transitions in the same element
is difficult. Using two different elements will introduce uncertainty as to the equality
of the two $Z$. Finally, computing the epsilon expansion coefficients for elements other than
hydrogen involves a considerable effort, with greater error in the theoretical $a$ and
$b$ values compared to hydrogen.

Using standard error propagation, the error associated with
$\epsilon$, $\delta \epsilon$, can be calculated. Thus

\begin{equation}
\delta \epsilon^{2}=\left\{ \left[-{1\over a}+{\Lao\over
a^{2} \left(\dLo\right)}\right]\delta \Lal\right\} ^{2} +
\left\{ \left[{\Lao\over ab\left(\dLo\right)}\right]\delta
\Lbl\right\} ^{2} $$
$$+ \left\{ \left[-{\Lao\left(\dLl\right)\over
a^{2}\left(\dLo\right)^{2}}+{\left(\dLl\right)\over
a\left(\dLo\right)}\right]\sigma _{\Lao}\right\} ^{2}$$
$$+\left\{
\left[{\Lao\left(\dLl\right)\over ab\left(\dLo\right)^{2}}\right]\sigma
_{\Lbo}\right\} ^{2}$$
$$+ \left\{ \left[{\Lal\over
a^{2}}+{{\Lao}^{2}\left(\dLl\right)\over
a^{3}\left(\dLo\right)^{2}}-{\Lal\Lao\over
a^{3}\left(\dLo\right)}-{\Lao\left(\dLl\right)\over
a^{2}\left(\dLo\right)}\right]\delta a\right\} ^{2}$$
$$+ \left\{
\left[-{\Lao\left(\dLl\right)\Lbo\over
ab^{2}\left(\dLo\right)^2}+{\Lao\Lbl\over
ab^{2}\left(\dLo\right)}\right]\delta b\right\} ^{2}
\end{equation}

Here $\delta\lambda^{lab}_{\alpha}$ and 
$\delta\lambda^{lab}_{\beta}$ are the uncertainties in the lab 
(present epoch) wavelengths (taken to be $0.01 \AA$, since we 
average over the hyperfine Lyman doublet), 
$\sigma_{\lambda_{\alpha}^{obs}}$ and 
$\sigma_{\lambda_{\beta}^{obs}}$ are the standard deviations of 
the measured red-shifted lines, and $\delta a, \delta b$ are the 
theoretical uncertainties in the epsilon expansion values (taken 
as $\delta a/a = \alpha = \delta b/b$). This equation includes 
all the errors associated with the quantities determining the 
value of $\epsilon$, for each matching Ly$_{\alpha}$ and 
Ly$_{\beta}$ data pair. For example, the absorption widths of 
Ly$_{\beta}$ are larger than Ly$_{\alpha}$ due to contamination 
by lower redshift Ly$_{\alpha}$ lines. Thus the centroid of the 
line has an ambiguity, resulting in a possible non-zero 
$\epsilon$ for that pair, having nothing to do with $D \neq 3$. 
This spectral contamination leads to an increase in the line 
width, so even though $\epsilon$ picks up a non-zero 
contribution, so too does $\delta \epsilon$, and it is the 
relative magnitude of these two quantities which determines the 
significance of the data pair.

The SIMBAD database was searched for papers having either emission or absorption
spectra containing matching Ly$_{\alpha}$ and Ly$_{\beta}$ lines. Those
spectra which have multiple very closely grouped Ly$_{\alpha}$ and Ly$_{\beta}$
lines, indicating the presence of closely spaced hydrogen cloud groupings,
were not used due to the possible ambiguity of assigning which Ly$_{\alpha}$
line goes with which Ly$_{\beta}$ line. This non-ambiguity filter
and the fact that of the many references identified in the database, very few
contained tabular identification of both Ly$_{\alpha}$ and Ly$_{\beta}$ lines,
restricted
the number of matching pairs to 11, given in Table 2. In all cases, the original
authors (astronomers) identified the Lyman lines.

\begin{table}
\begin{tabular}{lllcllc}
QSO & $\lambda^{obs}_{\alpha}$ (\AA) & $\lambda^{obs}_{\beta}$
(\AA) & FWHM & $\sigma_{\alpha}$ (\AA) & $\sigma_{\beta}$ (\AA) &
ref \\ \hline Q1206+119 & 4890.79 & 4126.63 & 50 km/s & 0.3 & 0.2
& \cite{AD} \\ Q1315+472 & 4342.76 & 3664.28 & 50 km/s & 0.3 & 0.2 &
\cite{AD} \\ Q1315+472 & 4322.87 & 3647.27 & 50 km/s & 0.3 & 0.2 & \cite{AD}
\\ Q1315+472 & 4220.69 & 3561.08 & 50 km/s & 0.3 & 0.2 & \cite{AD} \\
Q1623+268 & 4289.70 & 3619.42 & 50 km/s & 0.3 & 0.2 & \cite{AD} \\
BR 0401-1711 & 6383.5 & 5396.3 & 5 \AA & 1.8 & 1.8 & \cite{DT} \\
BRI 1108-0747 & 6000.5 & 5050.7 & 5 \AA & 1.8 & 1.8 & \cite{DT} \\
BRI 1114-0822 & 6755.5 & 5659.7 & 5 \AA & 1.8 & 1.8 & \cite{DT} \\
BR 1117-1329 & 6130.6 & 5111.2 & 5 \AA & 1.8 & 1.8 & \cite{DT} \\
Q0010+008 & 4963.7 & 4190.2 & 3.6 \AA & 1.3 & 1.3 & \cite{LJS} \\
Q0014+813 & 5253.29 & 4432.44 & 0.06 \AA & 0.02 & 0.04 & \cite{CJH}
\end{tabular}
\caption{QSOs with matching Ly$_{\alpha}$ and Ly$_{\beta}$. Here
$\lambda^{obs}_{\alpha}$ and $\lambda^{obs}_{\beta}$ are the
measured redshifted transition wavelengths, FWHM is the spectral
resolution of the data, and $\sigma_{\alpha}$ and $\sigma_{\beta}$
are the standard deviations.}
\end{table}

We then transferred each authors' data into Mathematica for
analysis, and $\epsilon$
was computed for each matched pair using (5). We used weighted averages based
on the quantum mechanical intensities for the laboratory wavelengths
$\lambda_{\alpha}^{lab}$, $\lambda_{\beta}^{lab}$ from the (1/2-1/2) and
(1/2-3/2) transitions.
Next, we computed the uncertainty for
each $\epsilon$, $\delta \epsilon$, based on the listed spectral resolution from each
reference and the order($\alpha$) uncertainty due to higher order quantum corrections.
Table 3 lists these results.

\begin{table}
\begin{tabular}{llll}
QSO & $<Z>$ & \phantom{aaa}$\epsilon$ & \phantom{aa}$\delta
\epsilon$ \\ \hline 
Q1206+119 & 3.0222 & -0.000059 & 0.00095 \\
Q1315+472 & 2.5723 & -0.00023 & 0.0011 \\
Q1315+472 & 2.556 & \phantom{ }0.00052 & 0.0011 \\
Q1315+472 & 2.48 & \phantom{ }0.00045 & 0.0011 \\
Q1623+268 & 2.5287 & \phantom{ }0.000064 & 0.0011 \\
BR 0401-1711 & 4.236 & -0.022 & 0.0058 \\
BRI 1108-0747 & 3.922 & \phantom{ }0.030 & 0.0074 \\
BRI 1114-0822 & 4.495 & \phantom{ }0.094 & 0.016 \\
BR 1117-1329 & 3.958 & \phantom{ }0.17 & 0.030 \\
Q0010+008 & 3.076 & -0.0059 & 0.0049 \\
Q0014+813 & 3.32 & \phantom{ }0.00008 & 0.00019
\end{tabular}
\caption{$\epsilon$ with its uncertainty $\delta 
\epsilon$ for each mean QSO $Z$.} 
\end{table}

These data have a mean for $\epsilon$ of 0.02436, with a standard
deviation of 0.057. Figure 1 shows $\epsilon$ for each of the
emission and absorption line pairs, against the mean QSO $Z$
value for the pair. We next employed
Student's t-test (10 degrees of freedom) to check the null
hypothesis $\epsilon$ = 0. Computing
the value for t gave 1.4056. Statistically, this means that to the level of
significance of 0.05, we cannot reject $\epsilon$ = 0, but to the level
of significance of 0.10, it can be rejected \cite{GKK}. Finally a $\chi^{2}$ fitting
gave $\epsilon$ = 0 with a goodness-of-fit probability of 0.2867.

Concerning the data, \cite{CJH} ($Z$ = 3.32)
taken with the Keck 10m instrument,
is the highest quality.
The uncertainty in $\epsilon$, $\delta\epsilon$, due to the error
measurements in the red-shifted lines is considerably smaller
than epsilon itself for the data of reference \cite{DT}. This can be
an indicator that $\epsilon$ really is non-zero for high $Z \geq
4$ objects, or that the $Z \geq 4$ data are contaminated with some 
undetermined systematic error. 

\vspace{0.5cm}
\begin{figure}[htb]
\begin{center}
\leavevmode {\epsfysize=10cm \epsffile{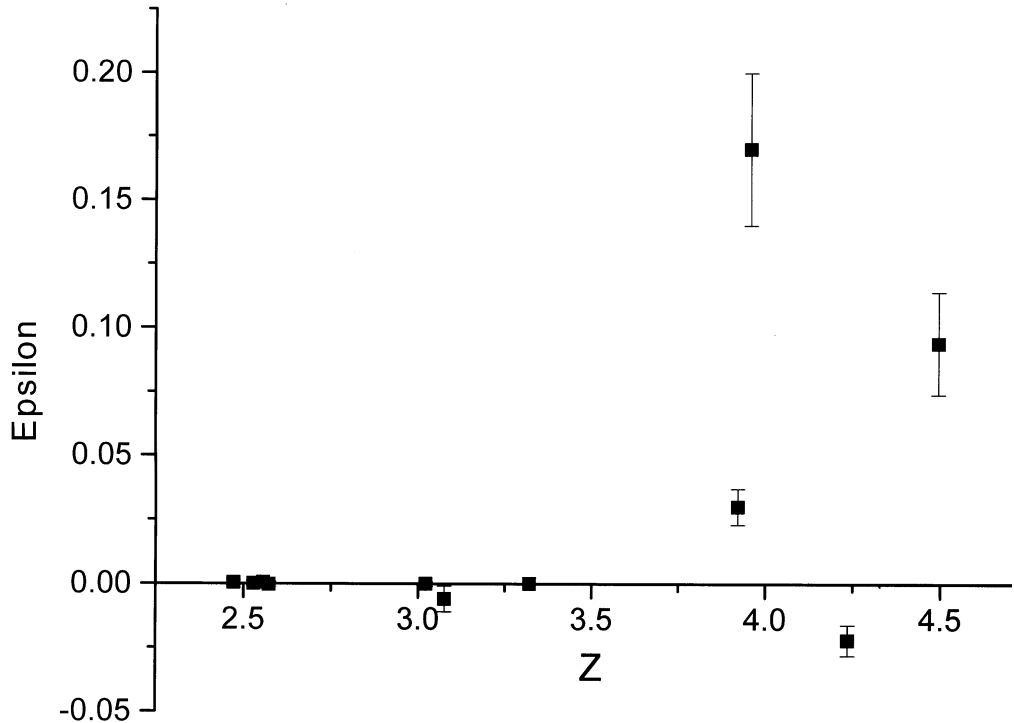}}
\end{center}
\caption[*]{\baselineskip 13pt
$\epsilon$ for each of the emission and absorption line
pairs, against the mean $Z$ value for the QSO.}
\label{usmth83f1} \end{figure}

In conclusion, the statistical tests and experimental data are
all consistent with $\epsilon = 0$ for $Z \leq 3.3$, but the 
experimental data support non zero $\epsilon$ values for $Z \geq 
4$. High spectral resolution data for $Z \geq 4$ would allow 
$\epsilon$ in the early universe to be better determined.

{\bf Acknowledgments: }
This research has made use of the SIMBAD database, operated at CDS,
Strasbourg, France. We would also like to thank the SIMBAD staff for their
assistance, and for providing one of us (DJB) with an account to
search their database for high $Z$ QSOs.

\end{document}